# The Vacancy Effect on Thermal Interface Resistance between Aluminum and Silicon by Molecular Dynamics

Yingying Zhang[1#], Xin Qian[1#], Zhan Peng[1], Nuo Yang[1*]

[1]School of Energy and Power Engineering, Huazhong University of Science and Technology (HUST), Wuhan 430074, People's Republic of China

**ABSTRACT**

Thermal transport across interfaces is an important issue for microelectronics, photonics, and thermoelectric devices and has been studied both experimentally and theoretically in the past. In this paper, thermal interface resistance (1/G) between aluminum and silicon with nanoscale vacancies was calculated using non-equilibrium molecular dynamics (NEMD). Both phonon-phonon coupling and electron-phonon coupling are considered in calculations. The results showed that thermal interface resistance increased largely due to vacancies. The effect of both the size and the type of vacancies is studied and compared. And an obvious difference is found for structures with different type/size vacancies.

**INTRODUCTION**

Interfacial thermal transport plays an important role for microelectronics, photonics, and thermoelectric devices where the devices reach nanoscale and have high interface densities[1, 2]. Recently, there are studies on the interfacial issue in experiment, theory and simulation. Since the acoustic mismatch model (AMM) and diffusive mismatch model (DMM) are limited in calculating the thermal conductance of ideal interface in cryogenic temperature[1, 2], the molecular dynamics (MD) simulation has become a popular method of predicting thermal boundary resistance in the last decade.

The advantages of MD are that there is no assumption on phonon transmission mechanics and it includes the anharmonic effects and inelastic scatterings[3-5]. There are some calculations of the interfacial thermal conductance between semiconductor and metal. Cruz *et al.* calculated the thermal interface conductance of Au/Si at 300 K[6], and the value of 188 MW/m$^2$-K is in good agreement with the measurement results with values ranging between 133 and 182 MW/m$^2$-K[7]. Wang *et al.* calculated interfacial thermal conductance of Cu/Si with values around 400 MW/m$^2$-K[8], which is much higher than experimental data[9].

Instead of an ideal interface, our recent work simulate a relaxed Al/Si interface with atomic level disorder by MD[10]. The value of thermal conductance is a little bit larger than the measurements. Interestingly, our results show that the localized ultra-high phonon modes, named as interface modes, emerge in disordered interface nanoscale region. That is, there is a novel and complex mechanism in the interfacial thermal transport.

More than two decades before, Swartz [1] predicted that the interface roughness would have effect on interfacial thermal transport. Recently, some experiments have shown the interface roughness, chemistry and structure could have a significant effect on thermal boundary conductance. Collins *et. al.* [11] presented experimentally that the surface chemistry affected the conductance between aluminum and diamond largely. Hopkins *et al*. [12] discovered that the increasing roughness and striping of the oxide layer at Si before deposition of Al would result in larger thermal boundary resistance. It was also shown that, experimentally, the thermal boundary conductance between quantum dot layers ($Ge_xSi_{1-x}$) and substrates (Si) could be controlled through the roughening of quantum dots[13].

In this work, using non-equilibrium MD (NEMD) simulation, we investigated the thermal transport across an Al/Si interface with vacancies. Three types of vacancies have been studied, and their effects on the thermal interface conduction are compared with each other.

**THEORY**

As shown in figure 1, the structure of Al/Si is along [100] direction. The interface distance is set as 0.26nm corresponding the minimum interface energy density. In order to minimize the interface mismatch, the cross section selects that Al 8×8 pairs up with Si 6×6 unit cells$^2$ (10.6nm$^2$). To simulate the vacancy, some atoms close to the interface are removed, which left a cylindrical hollow with height δ. The structures with different removed area are set for simulation. The vacancy ratio (R) is defined as the number of atoms removed over the atoms in the cuboid with the same height as δ.

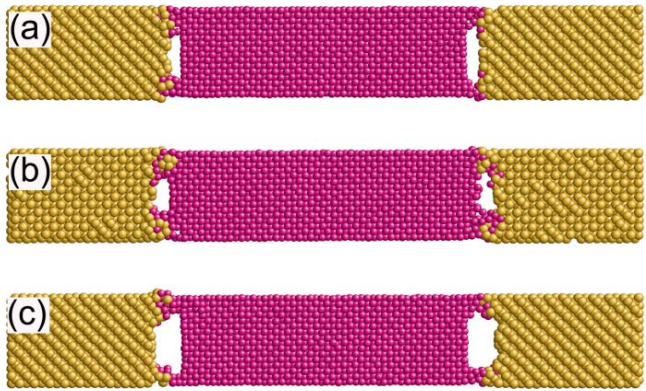

**Figure 1.** The sideview of simulation cell, Al 8×8×28 unit cells$^3$ /Si 6×6×21 unit cells$^3$ , after the relaxing process. The lattice constants of Al and Si are 0.4073 and 0.5431nm, respectively. The cross section is 10.6nm$^2$. As the vacancy is inside, to show it clearly, only a slice with a few layer atoms is plotted. (a) Structure with vacancy in Al at interface, where the vacancy ratio (R) is 0.316 and vacancy height (δ) is 0.407nm; (b) Structure with vacancy in Si, where R is 0.299 and δ is 0.407nm; (c) Structure with vacancy in both Al and Si, where R is 0.285 and δ is 0.815nm.

The second nearest-neighbor modified embedded atom method (2NN MEAM) interatomic potential, was adopted to describe the atomic interactions. The velocity Verlet algorithm is used to integrate the discretized differential equations of motions. The simulation time step, Δt, is chosen as 1 fs and the total simulation time is close to 3 ns. Before calculating the thermal interface conductance, the structure was fully relaxed. The structures after the relaxing process are shown in figure 1. (More details about MD simulation can be found in Ref.[10])

The interface thermal conductance of phonon ($G_{pp}$) and heat flux (J) are calculated as

$$G_{pp} = J/(A \cdot \Delta T) \quad (1)$$

Where ΔT is the interface temperature difference and A is the cross section area. The heat transferred across the interface can be calculated from

$$J = \frac{1}{N_t} \sum_{i=1}^{N_t} \frac{\Delta \varepsilon_i}{2\Delta t} \quad (2)$$

Where Δε is the energy added to/removed from each heat bath for each time step Δt. Both time and ensemble averaging is implemented to obtain a good statistics.

As proposed by Majumdar and Reddy[14], the resistance of electron-phonon in the metal and the resistance of interfacial phonon-phonon are in series,

$$1/G = 1/G_{ep} + 1/G_{pp} \quad (3)$$

where the interface conductance by electron-phonon coupling ($G_{ep}$) is calculated by the electron-phonon coupling constant ($g_{Al}$) and the phonon thermal conductivity of Al ($\kappa_{ph,Al}$) as $G_{ep} = \sqrt{g_{Al}\kappa_{ph,Al}} = 2.0$ GWm$^{-2}$K$^{-1}$. With the $G_{pp}$ from MD simulation, the total thermal interface conductance G can be calculate by Eq. (3). As shown in our previous work[10], the value of G is coincident with the value obtained from directly considering electron-phonon coupling in MD.

**DISCUSSION**

In the simulation, a temperature gradient is generated along the longitudinal direction normal to the interface by introducing heat baths at two sides of the interface with temperature $T_L$ and $T_H$ using Nosé-Hoover thermostat. In figure 2, it shows a typical time-averaged temperature profile and the definition of interface temperature difference ($\Delta T_{fit}$). In the inner regions, the temperature profile is linearly fitted. Near the interface, the temperature jumps nonlinearly due to the disordered atomic structure and the interface vacancy.

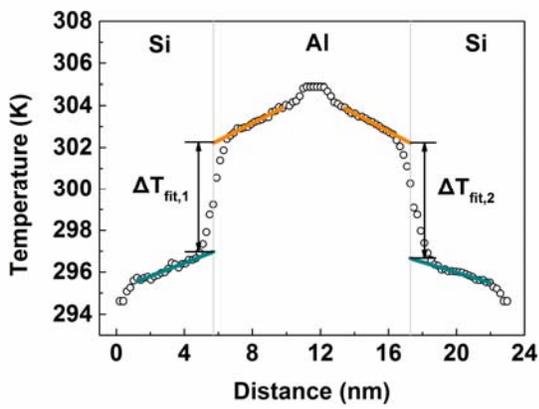

**Figure 2.** A typical time-averaged temperature profile of interface structure. The interface temperature difference is defined as difference between the linearly extrapolated temperature at each side of the interface. $\Delta T_{fit}=(\Delta T_{fit,1}+\Delta T_{fit,2})/2$.

### Size dependent

In our previous work on Al/Si without vacancy[10], there is a linear relationship between thermal interface resistance(1/G) and the inverse of system length(1/L), which means the extrapolation method can be used to calculate bulk thermal interface resistance. However, in the structure of Al/Si with vacancy, the thermal interface resistance is independent on the length (shown in figure 3). The value of bulk thermal interface resistance is obtained by averaging 1/G with different length, without using extrapolation method.

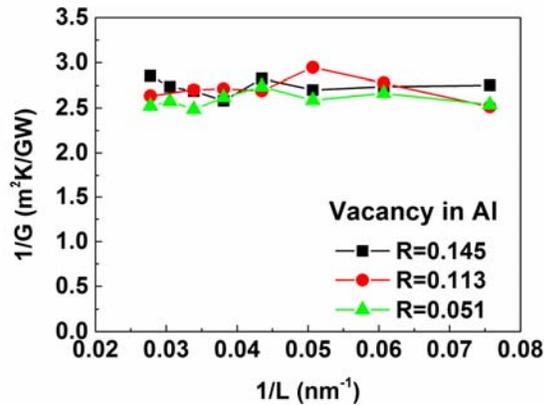

**Figure 3.** The independence of thermal interface resistance (1/G) on the length (L), for a structure shown in figure 1 (a), where the vacancy ratio (R) is 0.316 and vacancy height (δ) is 0.407nm.

In Al/Si without vacancy, there are more phonon eigenmode in a larger structure due to confinement effects in nanostructures. More phonon eigenmode takes more scatterings at interface, which makes a size dependence of G. However, in Al/Si without vacancy, there are more phonon boundary scatterings and more scatterings at the rough interface. The saturated phonon scatterings in the Al/Si with vacancy make value of G insensitive to the increase of size.

### Vacancy effect

To study the effect of vacancy, we calculated the thermal interface resistance of structures with several different vacancy ratios. As shown in figure 4, it is found that, with the increasing of vacancy ratio, the thermal interface resistance becomes larger. It is reasonable that a larger vacancy corresponds to a smaller contact area between Al and Si and more phonon boundary scatterings.

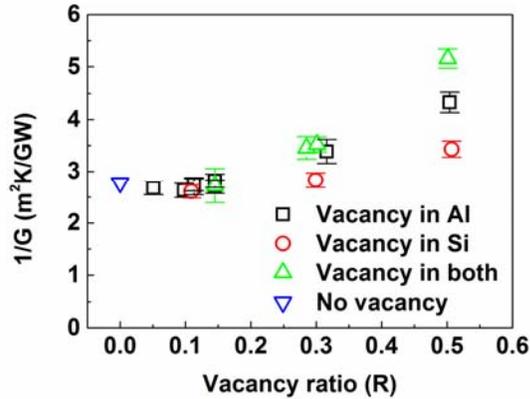

**Figure 4.** Thermal interface resistance (1/G) as a function of vacancy ratio R, for structures with three types of vacancies. The value of no vacancy refers to our previous work[10].

As shown is figure1, we also studied three types of vacancies: the vacancy in Al, in Si and in both Al and Si. The vacancy height in Al ($\delta_{Al}$) is 0.407nm, which is as the same as $\delta_{Si}$. $\delta_{Al+Si}$ is twice of $\delta_{Al}$, as 0.815nm. For the same contact areas, correspond to same R, there are an obvious larger resistance for structure with vacancy in both Al and Si than other two structures, especially for R above 0.2. For example, when R is 0.5, the value of 1/G for structure with vacancy in both Si and Al is as 1.25 times as that of vacancy in Al and as 1.5 times as that of vacancy in Si.

It is easy to be understood that the vacancy height can be looked as two times of rms roughness which will make more phonon scatterings at interface[13]. For R below 0.2, there is no obvious difference in resistance of different structures, and the values are close to experimental data(~2.857 m$^2$K/GW)[15].

For structures with vacancy in Al and with vacancy in Si, there is the same vacancy ratio and vacancy height. Interestingly, there is also a difference between them, especially for R above 0.2. A larger resistance for structure with vacancy in Al is shown in figure 4. It is found that there is difference in the contact area after the relaxing process, although the area is the same at the beginning of relaxation. The aluminum is softer and has a lower melting temperature than silicon. In the relaxation process of structure with vacancy in Si, some Al atoms at the interface would move into the vacancy of Si to increase the contact area. That is corresponding to smaller thermal interface resistance.

**CONCLUSIONS**

Using NEMD simulations, we show the calculation of thermal interface resistance of Al/Si structure with vacancies. The value of 1/G depends not only on the vacancy height, but the type of vacancy. A larger vacancy height will take more phonon scatterings which corresponding to a larger resistance. A lager contact area for structure with vacancy in Al would make a smaller the thermal interface resistance than with vacancy in Si.

## ACKNOWLEDGMENTS

N.Y. was sponsored by the grants from the National Natural Science Foundation of China Grant (11204216), Talent Introduction Foundation (0124120053) of HUST and Self-Innovation Foundation (2014TS115) of HUST. The authors thank the National Supercomputing Center in Tianjin (NSCC-TJ) for providing help in computations.

## REFERENCES

[#] Y.Z. and X.Q. contributed equally to this work.

[*] Corresponding author: N.Y. (E-mail: nuo@hust.edu.cn and imyangnuo@gmail.com)

1.  E. T. Swartz and R. O. Pohl, *Rev. Mod. Phys.* **61** (3), 605-668 (1989).
2.  D. G. Cahill, P. V. Braun, G. Chen, D. R. Clarke, S. Fan, K. E. Goodson, P. Keblinski, W. P. King, G. D. Mahan, A. Majumdar, H. J. Maris, S. R. Phillpot, E. Pop and L. Shi, *Appl. Phys. Rev.* **1** (1), 011305 (2014).
3.  E. S. Landry and A. J. H. McGaughey, *Phys. Rev. B* **80** (16), 165304 (2009).
4.  Y. Chalopin, K. Esfarjani, A. Henry, S. Volz and G. Chen, *Phys. Rev. B* **85** (19), 195302 (2012).
5.  R. J. Stevens, L. V. Zhigilei and P. M. Norris, Int. *J. Heat Mass Tran.* **50**, 3977-3989 (2007).
6.  C. A. d. Cruz, P. Chantrenne and X. Kleber, *J. Heat Transf.* **134**, 062402 (2012).
7.  P. L. Komarov, M. G. Burzo, G. Kaytaz and P. E. Raad, *Microelectr. J.* **34** (12), 1115-1118 (2003).
8.  Y. Wang, X. Ruan and A. K. Roy, *Phys. Rev. B* **85** (20), 205311 (2012).
9.  J. Xu and T. S. Fisher, Int. *J. Heat Mass Tran.* **49**, 1658-1666 (2006).
10. N. Yang, T. Luo, K. Esfarjani, A. Henry, Z. Tian, J. Shiomi, Y. Chalopin, B. Li and G. Chen, *J. Compt. Theor. NanoSci. (to be published in 2015)*, arXiv: 1401.5550.
11. K. C. Collins, S. Chen and G. Chen, *Appl. Phys. Lett.* **97** (8), 083102 (2010).
12. P. E. Hopkins, L. M. Phinney, J. R. Serrano and T. E. Beechem, *Phys. Rev. B* **82** (8), 085307 (2010).
13. P. E. Hopkins, J. C. Duda, C. W. Petz and J. A. Floro, *Phys. Rev. B* **84**, 035438 (2011).
14. A. Majumdar and P. Reddy, *Appl. Phys. Lett.* **84** (23), 4768-4770 (2004).
15. A. J. Minnich, J. A. Johnson, A. J. Schmidt, K. Esfarjani, M. S. Dresselhaus, K. A. Nelson and G. Chen, *Phys. Rev. Lett.* **107**, 095901 (2011).